\documentclass{article}
\usepackage[margin=2cm]{geometry}
\usepackage{graphicx} 
\graphicspath{{}}
\usepackage{hyperref}
\usepackage{siunitx}
\usepackage{booktabs}
\usepackage{makecell}
\usepackage[table]{xcolor}
\usepackage{amsmath}
\usepackage{amssymb}
\usepackage{subcaption}
\usepackage{authblk}
\usepackage[section]{placeins}
\usepackage[mathlines]{lineno}
\usepackage{setspace}
\title{JulianA: An automatic treatment planning platform for intensity-modulated proton therapy and its application to intra- and extracerebral neoplasms}
\date{May 2023}

\newbox{\myorcidaffilbox}
\sbox{\myorcidaffilbox}{\large\includegraphics[height=1.7ex]{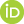}}
\newcommand{\orcidaffil}[1]{%
  \href{https://orcid.org/#1}{\usebox{\myorcidaffilbox}\,#1}}

\author[1]{Renato Bellotti \orcidaffil{0000-0002-2702-0437}}
\author[1]{Jonas Willmann \orcidaffil{0000-0002-2531-3551}}
\author[1]{Antony J.\ Lomax}
\author[2]{Andreas Adelmann \orcidaffil{0000-0002-7230-7007}}
\author[1,3,4]{Damien C.\ Weber \orcidaffil{0000-0003-1166-8236}}
\author[1,*]{Jan Hrbacek}

\affil[1]{Center for Proton Therapy, Paul Scherrer Institut}
\affil[2]{Accelerator Modelling and Advanced Simulations group, Paul Scherrer Institut}
\affil[3]{Department of Radiation Oncology, University Hospital of Zurich, USZ, Zurich}
\affil[4]{Department of Radiation Oncology, University Hospital of Bern, Inselspital, Bern}

\affil[*]{Corresponding author: jan.hrbacek@psi.ch, Forschungsstrasse 111, 5232 Villigen PSI, Switzerland }

\begin{document}

\maketitle

\begin{abstract}
    \textbf{Background}
    Creating high quality treatment plans is crucial for a successful radiotherapy treatment. However, it demands substantial effort and special training for dosimetrists. Automated treatment planning systems have been developed, but typically require either an explicit prioritization of planning objectives, human-assigned objective weights, large amounts of historic plans to train an artificial intelligence or long planning times. Many of the existing auto-planning tools are difficult to extend to new planning goals, which is a requirement for research and clinical usage.

    \textbf{Purpose}
    To develop a novel treatment planning platform for intensity-modulated proton therapy that is easy to use for research and creates clinical-grade treatment plans without human interaction.

    \textbf{Methods}
    A new spot weight optimisation algorithm, called JulianA, was developed. The name is a portmanteau between the name of our in-house treatment planning system FIonA and Julia, the programming language in which it is implemented. The algorithm minimises a scalar loss function using a gradient-based optimisation algorithm. The loss function is built only based on the prescribed dose to the tumour and organs at risk (OARs), but does not rely on historic plans. The objective weights in the loss function have default values that do not need to be changed for the patients in our dataset. However, human fine tuning is possible for especially challenging cases.
    The system was implemented in Julia, which has a script-like syntax, allows interactive development and has a runtime performance comparable to programming languages like C. Therefore, it is a versatile tool for researchers and clinicians without specialised programming skills. Extending the system is as easy as adding an additional term to the loss function. JulianA was validated on a dataset of $19$ patients with intra- and extracerebral neoplasms within the cranial region that had been treated at our institute. For each patient, a reference plan which was delivered to the cancer patient, was exported from our treatment database. Then JulianA created another treatment plan, called the auto plan, using the same beam arrangement. The reference and auto plans were given to a blinded independent reviewer who assessed the acceptability of each plan, ranked the plans and assigned the human-/machine-made labels.

    \textbf{Results}
    The auto plans were considered acceptable in $16$ out of $19$ patients. The reference plan was considered superior for $8$ cases, both plans were considered equivalent in $4$ cases and the auto plan was considered superior in $7$ cases. Whether a plan was crafted by a dosimetrist or JulianA was only recognised for $9$ cases. It cannot be shown that the reviewer has a better recognition rate than random guessing ($p \approx 1.0$). The median time for the spot weight optimisation is approx.\ $\SI{2}{min}$ (range: $\SI{0.5}{min} - \SI{7}{min}$).

    \textbf{Conclusions}
    A proof-of-concept evaluation of JulianA has been presented. JulianA provides competitive planning times, is easy to extend and does not need human interaction. The system creates treatment plans comparable to human-created plans for a majority of patients.
\end{abstract}
\clearpage

\section{Introduction}
Treatment planning is an important yet time consuming part of radiation therapy. High quality treatment plans maximise the tumour control probability while keeping the risk of radiation-induced adverse effects minimal. However, realising such a plan takes up to several hours in some challenging cases even by experienced treatment planners. Automatic treatment planning aims to alleviate this issue and bears several advantages. First, it reduces the general workload of human planners, allowing them to spend more time on especially challenging patients. Second, it sets a minimum plan quality standard and can therefore be used as a quality assurance tool. Third, it allows untrained researchers to incorporate realistic treatment plans into their research, which would otherwise be either impossible or impose additional workload onto clinical personnel. Finally, automatic treatment planning reduces interplanner variability of the plan quality.

Treatment planning solves two coupled problems such that the prescription, i.\ e.\ dose to the tumour and dose constraints to organs at risk (OARs), is fulfilled: the choice of a beam arrangement and the optimisation of the spot weights (or fluence map in the case of photons). This work will focus on proton therapy planning only, with an attention to spot weight optimisation (SWO) and assuming the beam arrangement is given. This is a reasonable assumption because the Center for Proton Therapy at the Paul Scherrer Institut has guidelines for indication-specific standard beam arrangements.

Several methods have been proposed to solve the SWO problem in an automatic or semi-automatic way. A non-exhaustive summary is given in Tab.\ \ref{tab:literature_summary}. The $2p \epsilon c$ algorithm \cite{2pepsilonc} receives a patient-unspecific wish list that prioritises the planning objectives and then optimises the highest-priority objective. Once the optimisation is converged, the achieved objective value is used as a constraint while optimising the second priority objective and so on. This results in a Pareto-optimal plan, i.\ e.\ no objective can be improved without worsening any other objectives, but requires an institution-wide consensus on the priority of objectives. Additionally, some objectives rely on biological parameters that need to be selected. The LRPM method \cite{lrpm}, \cite{lrpm2} is an extension of the $2p \epsilon c$ method that removes the need for strict prioritisation. It uses human assigned reference points and tradeoff parameters to guide the optimisation algorithm and contains biology-inspired terms.

The ECHO algorithm \cite{echo} solves a series of linear programming problems using a commercial solver, which turns the optimisation process into an uninterpretable blackbox and leads to high run times. On the other hand, the process is fully automatic and requires no input beyond the prescription.

Sayed et al.\ \cite{sayed} proposed to use differential evolution \cite{DE} in combination with Monte Carlo dose calculations to approximate the entire Pareto front of treatment plans. This approach presents all possible tradeoffs to a human decision maker who then has to select the final plan. Therefore, the algorithm is not fully automatic, but very flexible. However, it requires several hours even on a system with several graphics processing units (GPUs).

The POPS algorithm \cite{pops} uses projections onto the Pareto front and a Nelder-Mead simplex search to search clinically acceptable plans on the Pareto front, but needs more than an hour to create the treatment plan.

The BCQ-ARM algorithm \cite{projection} assigns an upper and lower dose bound to each voxel. Based on these voxel-wise constraints, a feasible solution is searched using the automatic relaxation method \cite{automatic_relaxation_method}. A bisection search on the bounds is performed as long as a feasible solution is found in order to optimise the objectives. This is performed for each OAR separately and the resulting spot weight vectors are summed up, weighted by a human-assigned importance. Default importances are provided. This method is not directly interpretable because the spot weight vectors are weighted, but not the objectives. 

The open source software matRad provides a toolkit to the users based on which they can build their own loss function as a weighted sum of various terms \cite{matrad}. The loss function is then minimised using the interior-point method IPOPT \cite{ipopt}. However, the weights of each loss term need to be set manually, which renders matRad non-automatic. On top, matRad is built on the proprietary Matlab platform, which increases the cost for research and clinical operation. However, it must be noted that matRad can be used on the free and open source platform Octave~\cite{octave}, albeit without graphical user interface.

Bangert et al.\ \cite{ultra_fast} suggest to optimise a weighted sum of one-sided quadratic terms. Each term punishes underdosage or overdosage to a specific voxel. The loss function is convex and the implementation tailored towards its specific formula, which renders the optimisation extremely fast, but impossible to extend without losing performance. The loss function is optimised using the L-BFGS algorithm \cite{lbfgs}. The authors do not mention how they obtain the voxel bounds, therefore we assume they are human input.

Guo et al.\ propose a weighted sum loss function similar to the ones matRad provides. However, they minimise the loss function using a gradient-based method and propose a heuristic way to update the importance of each term during the optimisation in order to align it with the clinical intent. Each term has its own update rule. The dynamic nature of the loss function removes not only the need for human interaction, but also the interpretability from the optimisation process. Further, adding additional terms requires also adding a new update rule for the corresponding objective weight, which is not trivial.

Knowledge-based planning methods make use of historic treatment plans to predict good planning parameters. Various approaches exist, but listing them is beyond the scope of this work. The interested reader is referred to review papers \cite{data_driven}, \cite{data_driven2}, \cite{data_driven3}. KBP methods share some common properties: they rely on training data whose performance they cannot exceed, they are typically non-interpretable and extending the model is impossible without retraining it. We do not consider KBP further due to their reliance on high quality consistent datasets, which are not available at every institute.

All of the existing methods have strengths and limitations. Most of them were not designed with extensibility and/or interpretability in mind. However, these properties are important in research. Some methods deliver Pareto-optimal plans, but take a long time to complete.

\begin{table}
    \rowcolors{2}{gray!25}{white}
    \centering
    \resizebox{0.9\textwidth}{!}{%
        \begin{tabular}{rrrrrrr}
            \toprule
            Method & Human input & Interpretable? & Extensible & Pareto optimal & automatic & Runtime\\
            \midrule
            $2p \epsilon c$ \cite{2pepsilonc} & \makecell[rt]{Wish list\\Biological parameters} & yes & yes (I) & yes & yes & -\\
            LRPM \cite{lrpm}, \cite{lrpm2} & \makecell[rt]{prescription\\Biological parameters} & yes & yes & yes & yes & 0.5 - 14.5min\\
            ECHO \cite{echo} & prescription & \makecell[rt]{no} & yes (I) & yes & yes & 20-170min\\
            DE \cite{sayed} & prescription & yes & yes (I) & yes & semi & \makecell[rt]{3.8 - 5.7h\\(multi-GPU)}\\
            POPS \cite{pops} & prescription & yes & yes (I) & yes& yes & \SI{78}{min}\\
            BCQ-ARM \cite{projection} & \makecell[rt]{prescription\\objective importance (optional)} & no & no & no & yes & 3 - 14min (CPU)\\
            Matrad \cite{matrad} & \makecell[rt]{prescription\\objective importances} & yes & yes & no & no & 0.5 - 8min (CPU)\\
            Bangert et al.\ \cite{ultra_fast} & \makecell[rt]{prescription\\lower\&upper dose per voxel} & no & partially & no & yes & 0.5 - 4s (CPU)\\
            Guo et al.\ \cite{factor_update} & prescription & yes & yes & no & yes & 3-11min\\
            KBP methods & Training data & no & yes (I) & no & yes & seconds\\
            JulianA (this work) & \makecell[rt]{prescriptions\\objective importances (optional)} & yes & yes & no & yes & 0.5 - 7min (GPU)\\
            \bottomrule
        \end{tabular}
    }
    \caption{Comparison of existing optimisation methods for spot weight/fluence map optimisation. The expression "yes(I)" means that extending the algorithm with additional goals is possible, but will increase its runtime. Notice that multiple methods produce Pareto optimal plans, but the objectives they are optimising are generally not the same. "Interpretable" means that the convergence can be analysed to infer the bottle neck and reasoning of the algorithm in high level concepts (e.\ g.\ OAR constraint violation for a specific OAR, underdosage, overdosage of the target, ...). Extensibility refers to extensibility w.\ r.\ t.\ arbitrary, especially non-linear and/or non-convex, objectives.}
    \label{tab:literature_summary}
\end{table}

We present JulianA, which is a novel auto planning algorithm using gradient-based optimisation of a loss function. The loss function follows the path of the well-established weighted sum approach \cite{Oelfke2001}. Contrary to existing methods, it does not only include terms to punish exceeding maximum and minimum dose thresholds for each voxel, but also incorporates physical prior knowledge as used in the original IMPT paper~\cite{Lomax_1999} and takes OAR sparing into account as high-level concepts. Further, a novel term for ensuring dose homogeneity within the tumour is used. This combination of prior knowledge with a homogeneity term and a high-level formulation of OAR sparing has, to the best of our knowledge not yet been applied. The advantages of the weighted sum loss are its flexibility and extensibility to future planning goals. Further, it can be optimised efficiently using gradient-based optimisation algorithms like L-BFGS~\cite{lbfgs}, making it useable and practical for both research and clinical adoption.

JulianA was implemented in a novel platform for treatment planning research at our institute, which bears the same name. It was developed for two main use cases: Patient selection and increase of productivity in research. JulianA can assist patient selection by generating a treatment plan of similar quality than the plans generated by dosimetrist in a short time, without increasing the workload to clinical personnel. On the other hand, it can increase research efficiency because many researchers have a project of which the treatment planning is only a small part. However, the lack of training and experience in treatment planning results in low productivity and potentially unrealistic treatment plans, which limits the transfer of new insight and methods into the clinic. To alleviate this problem, our new method relies on minimum human input, namely only the prescription. It is easily extensible by domain experts without experience in treatment planning. In order to keep research productivity high and the entry barrier low, the platform is implemented in Julia, which is a high-level and open source programming language. Julia is flexible and versatile, produces fast code that can utilise modern computing hardware such as GPUs and the latest developments in computer science, e.\ g.\ for optimisation, without much effort. This removes the need to use different programming languages for research and production environments, which is known as the two language problem. Julia comes with interactive execution and debugging capabilities, which further improves research productivity. Having a scriptable treatment planning pipeline makes research more easily reproducible, too. We hope that JulianA will also boost the productivity in our clinic. It can be used as a quality assurance tool by providing a minimum plan quality standard and reduces the workload on human planners. The long-term clinical goal of this platform is to reduce the time needed for patient selection by providing a quick estimate of the potential of proton radiotherapy. Such a plan could then be compared to an automatically generated photon plan in a similar way to the one proposed by Langendijk et al.~\cite{LANGENDIJK2013267}. We validate JulianA in a blind study with a radiation oncologist.

\section{Methods}

\subsection{New spot optimisation algorithm}
Our new algorithm is optimising the plan by minimising $\mathcal{L}(\mathbf{d}) \in \mathbb{R}$, the scalar loss function 
show in Equ.~\ref{equ:loss1}
\begin{equation} \label{equ:loss1}
    \mathcal{L}(\mathbf{d}) = \alpha_{ideal} \mathcal{L}_{ideal} + \alpha_{hom} \mathcal{L}_{hom} + \alpha_{min} \mathcal{L}_{min} + \alpha_{max} \mathcal{L}_{max} + \sum_o \alpha_{oarmean} \mathcal{L}_{mean}(o) + \alpha_{oarmax} \mathcal{L}_{max}(o), 
\end{equation}
where $\mathbf{d} = D \cdot \mathbf{w} \in \mathbb{R}^N$ is the vector containing the dose at each of the $N$ optimisation point, $D$ is the influence matrix calculated by our in-house planning system FIonA and $\mathbf{w}$ is the spot weight vector. Optimising a plan means minimising $\mathcal{L}(\mathbf{d})$.

The loss function (Equ.~\ref{equ:loss1}) is a mathematical representation of the clinical intent summarized by 6 terms c.f.\ Equ. \ref{equ:ideal1} to Equ. \ref{equ:oar_max1}. It can easily be extended and the tradeoff parameters $\alpha_x$ allow a human planner to control the tradeoffs. The placeholder $x$ stands for {\tt ideal}, {\tt hom}, {\tt min}, {\tt max}, {\tt oarmean}, {\tt oarmax}. 
However, we found that the parameter values in Tab.~\ref{tab:tradeoff_weights} work very well for the diverse set of patients in our dataset (see following subsections). We did not have to adjust any of the parameters in a patient specific way, which renders our method for spot weight optimisation fully automatic. Future work is needed to investigate whether these values are also suitable for other datasets, especially other treatment sites, than the one investigated in this work.

Each of the 6 terms in the loss function 
represents a different planning goal. What terms are present for the OARs depends on the prescription, which is the only input to our algorithm. It is patient-specific and the same input a human planner would obtain. The prescription contains the prescribed dose $D_p$ to the target as well as mean and/or maximum dose constraints the OARs. A mean or maximum dose constraint for OAR $o$ is given by $\mu(o, \mathbf{d}) \leq \mu_{c}(o)$ and $D_{max}(\mathbf{d},o) \leq M_c(o)$, respectively, where $\mu(o, \mathbf{d})$ denotes the mean dose and $D_{max}(\mathbf{d},o)$ the maximum dose in OAR $o$ for dose distribution $\mathbf{d}$. Multiple targets with different prescribed dose levels may be present. JulianA can handle these situations, but for the sake of simplicity of notation we assume only one prescribed dose level. The terms in the loss function are given by
\begin{align}
    \mathcal{L}_{ideal} &= \sum_{i=1}^N I_i (d_i - d_{ideal, i})^2, \label{equ:ideal1}\\
    \mathcal{L}_{hom} &= \frac{1}{N_{target}} \sum_{i \in target} (d_i - \mu(target, \mathbf{d}))^2, \label{equ:hom1}\\
    \mathcal{L}_{min} &= \sum_{i=1}^N |d_{min, i} - d_i|_+, \label{equ:min1}\\
    \mathcal{L}_{max} &= \sum_{i=1}^N |d_i - 1.05 D_p|_+, \label{equ:max1}\\
    \mathcal{L}_{mean}(o) &= |\mu(o, \mathbf{d}) - \mu_{c}(o)|_+, \label{equ:oar_mean1}\\
    \mathcal{L}_{max}(o) &= \sum_{i \in o} \left| d_i - M_c(o) \right|_+, \label{equ:oar_max1}
\end{align}
where the sums go over the optimisation point indices $i$, i.\ e.\ the locations at which we probe the dose distribution. The function $|\cdot|_+ = \max(\cdot, 0)$ is the identity for positive values and zero for non-positive values and $N_{target}$ denotes the number of voxels that belong to the target structure.

Equation~\ref{equ:ideal1}, the first term in $\mathcal{L}(\mathbf{d})$, penalises deviations from an ideal dose distribution that is inspired by the physics of pencil beams. It is the weighted sum of squared differences between the current dose distribution $\mathbf{d}(\mathbf{w})$ and an ideal dose distribution introduced in the original paper for intensity-modulated proton therapy \cite{Lomax_1999}. This models the desirable dose distribution as a plateau at the prescribed dose level within the defined target volume, and with a Gaussian falloff outside the target. The ideal dose value at optimisation point $i$ and its importance are calculated according to
\begin{align*}
    d_{ideal, i} &= D_p \exp \left( \frac{-\Delta_i^2}{2 \sigma^2} \right)\\
    I_i &= \min\left(\frac{d_{ideal, i}}{f D_p}, 1\right)^\gamma,
\end{align*}
where $\Delta_i$ is the minimum in-slice distance of optimisation point $i$ from the target contours (zero for optimisation points inside the target) plus a falloff distance of \SI{1.672349}{cm}, the parameter $\sigma$ is the standard deviation of the Gaussian falloff outside the target volume, $f = 0.9$ is the dose at edge factor and $\gamma = 10$ is an importance factor. The values of these parameters have been in clinical use in our in-house planning system for several years. We adjust the ideal dose distribution to ensure that it does not violate maximum dose OAR constraints.

The term $\mathcal{L}_{hom}$ (Equ.~\ref{equ:hom1}) penalises variance within the targets, therefore incentivising a more homogeneous dose distribution within the target. With $\mathcal{L}_{min}$ (Equ.~\ref{equ:min1}) we aim to ensure a sufficient target coverage by introducing a minimum dose distribution. It can be calculated by
\begin{align*}
    d_{min, i} &= \begin{cases}
        d_{ideal, i} - \SI{0.5}{Gy}, &i \in \mathrm{target},\\
        0, &\mathrm{otherwise}
    \end{cases}.
\end{align*}
The minimum dose distribution $\mathbf{d}_{min}$ is modified to make sure its values are below all OAR maximum dose prescriptions for each optimisation point. With $\mathcal{L}_{max}$ (Equ.~\ref{equ:max1}) we  ensure that no optimisation point receives more dose than $105\%$ of the prescribed dose. Therefore, it prevents hot spots. The final two terms, Equ.~\ref{equ:oar_mean1} and Equ.~\ref{equ:oar_max1}, penalise violations of OAR constraints.

The novelty of the algorithm is that it does not only use the terms $\mathcal{L}_{min}, \mathcal{L}_{max}$, but also includes physical knowledge in the form of the $\mathcal{L}_{ideal}$. The $\mathcal{L}_{hom}$ term has never been used to the best of our knowledge. Further, the definition of $\mathbf{d}_{min}$ takes overlapping OARs into account, which is also a novelty to the best of our knowledge. Finally, $\mathcal{L}_{mean}(o)$ formalises the OAR mean dose constraints as a high-level concept, instead of applying just a fix maximimum dose threshold to each voxel.

\begin{table}
    \centering
    \begin{tabular}{rrrrrrr}
        \midrule\\
        Weight & $\alpha_{ideal}$ & $\alpha_{hom}$ & $\alpha_{min}$ & $\alpha_{max}$ & $\alpha_{oar, mean}$ & $\alpha_{oar, max}$\\
        Value & $\frac{1}{25}$ & $1$ & $\frac{1}{100}$ & $10$ & $1$ & $1$\\
        \bottomrule
    \end{tabular}
    \caption{Subloss weights and their values.}
    \label{tab:tradeoff_weights}
\end{table}

\subsection{Implementation}
The entire code is implemented in the Julia language \cite{julia} (version $1.8.1$), which has a script-like syntax but a runtime comparable to C thanks to the multiple dispatch paradigm and just-in-time compilation. The language was developed specifically for scientific programming and interactive development. It allows trivial usage of linear algebra on GPUs thanks to the \texttt{CUDA.jl} library\cite{cuda_jl}. There are also libraries for other GPU manufacturers than NVIDIA. The derivatives of the loss function are obtained using automatic differentiation with the \texttt{Zygote.jl} library \cite{zygote}, which allows a researcher to focus on writing meaningful loss functions without having to implement a numerical or analytical gradient calculation. Additionally, automatic differentiation allows to evaluate gradients pointwise even for functions for which no closed form solution exists. The loss function is optimised using the L-BFGS \cite{lbfgs} implementation in the \texttt{Optim.jl} Julia package \cite{optim.jl}. All spots are initialised to a uniform weight. In each iteration, the dose is renormalised before the loss is calculated. The optimisation stops when the loss value does not decrease by more than $0.5$ for $25$ iterations. A checkerboard pattern at CT resolution is applied around a margin of the target to select at which positions the dose should be calculated during the optimisation.

\subsection{Blinded expert evaluation of  plans}

An independent expert reviewer assessed the clinical acceptability of each proton therapy plan and compared auto and reference plans for each patient. Auto plan dose distributions were calculated using our in-house treatment planning system (FionA), which is used clinically on a routine basis, following the approach described in this manuscript, whereas the reference plans were calculated in the predecessor to FionA (PSIPlan) following the clinical guidelines for treatment planning valid at the time of treatment. In PSIPlan, FionA and JulianA the same dose calculation algorithm is used (Raycasting). In addition, the same field arrangements and spot and energy separation/placement strategies as used in the reference plans have been used for the auto-plans. Robust optimization has not been used for any of the plans, with all being planned directly to a PTV, as is standard clinical practice at our center for treatments on Gantry 2. The goal of this study is to provide a proof-of-concept for the JulianA system, but future work is needed to proof generalisation beyond a single reviewer. The expert reviewer was not involved in the treatment of the patients and was blinded to the source of the plan generation to minimize potential biases. The evaluation process involved the use of a standardised review form and criteria, as well as the expert reviewer's individual clinical judgement. The primary plan comparison metrics included target coverage, organ sparing, and high-dose conformity. The expert reviewer assessed the machine- and human-generated proton therapy plans individually and compared them on the following aspects: 1) clinical acceptability (yes or no); 2) superiority of the plan (yes or no); and 3) identification of the plan as human-generated (yes or no).

The dataset for the blinded evaluation consisted of $19$~patients with intra- and extracranial cancers that were treated between $2013-2020$ at gantry two of the Paul Scherrer Institut. Tab.~\ref{tab:patients} details each patient included in the analysis. All of the patients were treated in our institute. The dataset contains a diverse set of indications because the purpose of this proof-of-concept study is to explore the utlility of JulianA for a wide range of indications that are treated at our institute. The dataset contains the prescribed dose, the original planning CT, the contours of the targets and OARs and the dose distribution that was approved for treatment in our clinic. We call the latter the reference dose distribution. The JulianA spot weight optimisation algorithm was applied to each patient, which resulted in the auto plans and auto dose distributions. The same beam arrangement as for the reference plans was used where possible. In case the beam angles changed during the course of treatment of sequential radiation delivery, the beam arrangement that was used for the most fractions is the one selected for the autoplan. The planning target volumes (PTVs) were used as the targets. Their expansion as well as the prescribed doses and OAR constraints are patient specific and the same for both the reference and the auto plan. Simultaneous integrated boost plans were created for the auto plan if multiple dose levels were prescribed, even if the reference plan was generated as a multi-series plan. Currently JulianA can only generate SIB plans, but a dosimetrist could create series-specific prescriptions and constraints and run JulianA multiple times to create sequential plans. The reference and auto dose distributions were normalised in the same way to increase comparability. The mean dose to the PTV that receives the highest dose was scaled to be $100\%$ of the prescribed dose. The voxels belonging to an overlapping OAR were excluded unless such an exclusion would remove more than $50\%$ of the PTV voxels. 

\begin{table}[]
    \centering
    \begin{tabular}{rrr}
        \toprule
        Patient ID & Diagnosis and location & Prescribed Dose [Gy RBE] \\
        \midrule
        00 & Chordoma, clival & 73.8\\
        01 & Pilocytic astrocytoma, mesencephalic  & 50.4\\
        02 & Haemangioma, orbital & 20.0\\
        03 & Melanoma, conjunctival & 66.0\\
        04 & Meningioma, clival & 54.0\\
        05 & Meningioma, petroclival & 54.0\\
        06 & Pilomyxoid astrocytoma, optic nerve  & 52.2\\
        07 & Chordoma, clival & 74.0\\
        08 & Anaplastic astrocytoma, frontal  & 59.4\\
        09 & Chordoma, clival & 74.0\\
        10 & Germinoma, XXX & 24.0\\
        11 & Embryonal tumor with multilayered rosettes (ETMR), frontoparietal & 59.4\\
        12 & Embryonal rhabdomyosarcoma, intraorbital & 45.0\\
        13 & Atypical meningioma, XX & 60.0\\
        14 & Meningioma, optic nerve & 50.4\\
        15 & Chordoma, clival & 74.0\\
        16 & Ependymoma, prepontine & 59.4\\
        17 & Papillary ependymoma, thalamic & 59.4\\
        18 & Pilocystic astrocytoma, optical nerves & 54.0\\
        \bottomrule
    \end{tabular}
    \caption{Patients' characteristics (n=19) for the auto treatment planning analysis utilising JulianA} 
    \label{tab:patients}
\end{table}
\clearpage

\section{Results}
\begin{figure}
    \centering
    \includegraphics[width=\textwidth]{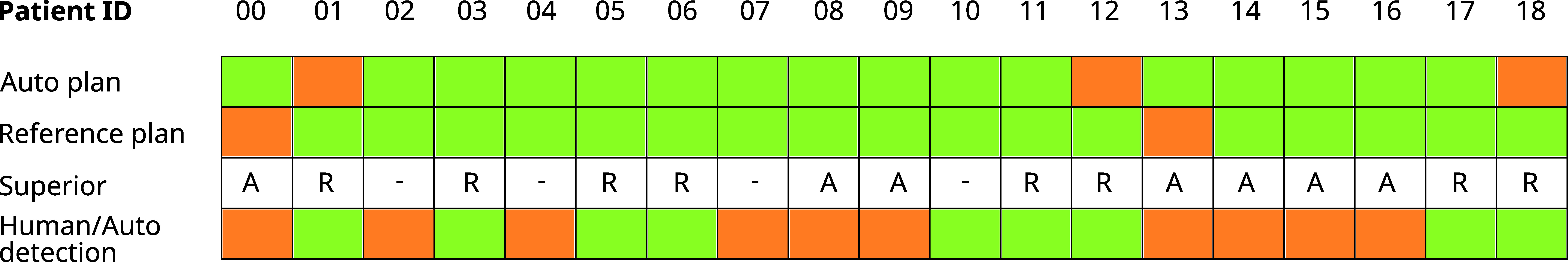}
    \caption{Results of the blind study. The first two lines indicate whether the plan was considered acceptable (green) or inacceptable (orange). The third row indicates which plan was considered superior ("A" means the auto plan, "R" means the reference plan and "-" means both plans are equivalent). The final row shows whether the physician recognises the reference plan and the auto plan correctly.}
    \label{fig:summary}
\end{figure}

The main findings of the blind study are summarised in Fig.~\ref{fig:summary}. We have observed that $16$ auto plans are acceptable for treatment without modification while $17$ of the reference plans were acceptable. These numbers are comparable. The fact that two reference plans were considered unacceptable might be due to missing information about the patient history and patient-specific planning goals beyond the prescription. The reference plan is considered superior for $8$ of the $19$ cases, equivalent for $4$ cases and inferior for $7$ cases. This implies that the automatic planning method arrives at a comparable or even better treatment plan for $11$ cases. Finally Fig.~\ref{fig:summary} shows for which patient the expert reviewer recognised the auto plan correctly, which can be seen as a variation of Alan Turing's famous test \cite{turingtest}. The physician was able to recognise the automatically generated plan in $9$ out of $19$ cases. The exact binomial test \cite{Conover1998} is performed for the null hypothesis that the physician's classification follows a binomial distribution with success probability $0.5$, i.\ e.\ the flip of a fair coin. The available data is not sufficient to reject the null hypothesis on the $5\%$ confidence level ($p \approx 1.0$).

A detailed table of dosimetric quantities is shown in Tab.~\ref{tab:metrics} and visualised in Fig.~\ref{fig:boxplots} in the appendix. The overall maximum dose (interquartile range and median of the difference $[0.4\%, 3.3\%]$, $1.4\%$ of the prescribed dose) is generally higher for the reference plan. The minimum dose to any target voxel is also higher for the reference plans ($[-1.0\%, 14.2\%]$, $9.7\%$). Given the comparable $V_{95\%}$ ($[-3.0\%, 3.0\%]$, $0.9\%$) and $HI = D_{5\%} - D_{95\%}$ ($[-1, 1.3]$, $-0.1$) performance, we assume that this indicates underdosage to a small number of voxels rather than a systematic deterioration of plan quality.

\begin{figure}
    \centering
    \begin{subfigure}{0.6\textwidth}
        \centering
        \includegraphics[width=\textwidth]{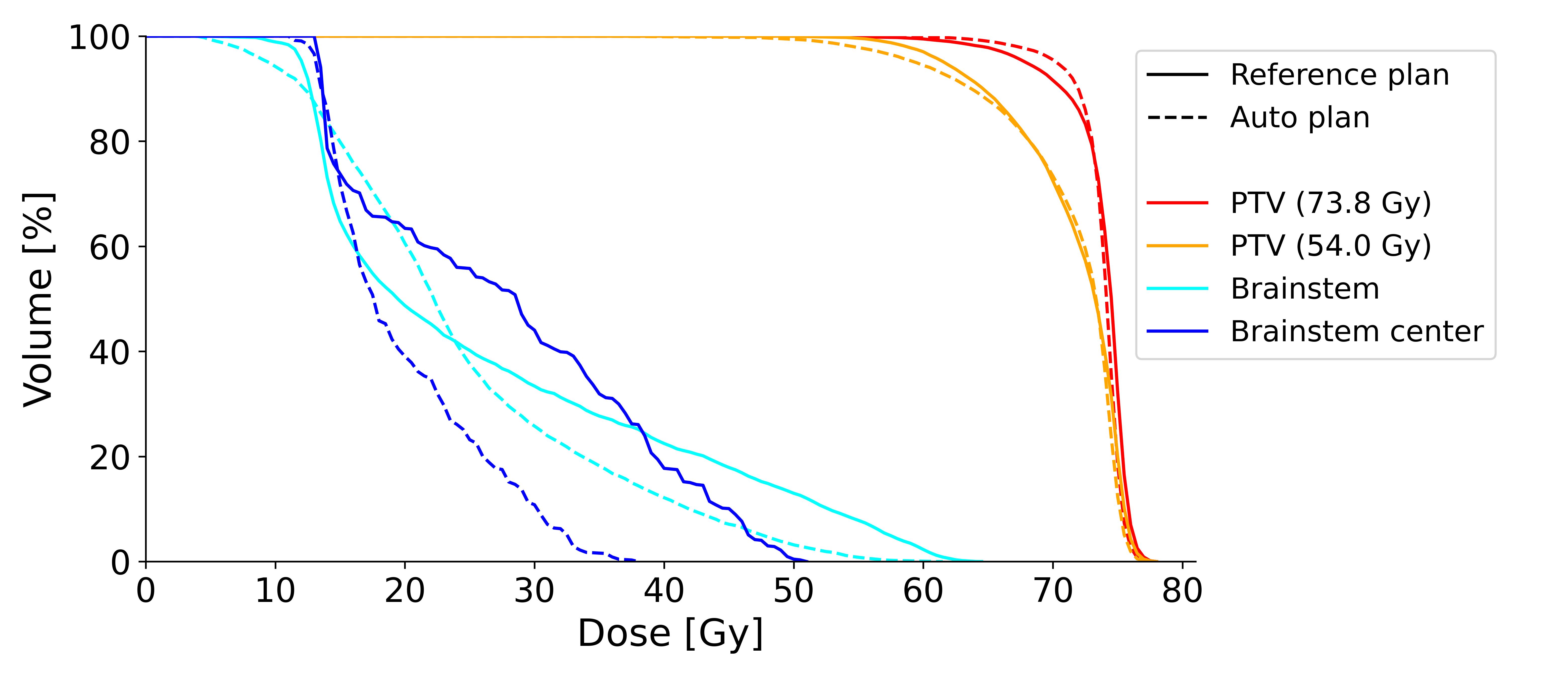}
        \caption{DVH curves for patient 00. The solid lines are the human-crafted reference plans, the dashed lines represent the auto plan.}
        \label{fig:00_dvh}
    \end{subfigure}\\
    \begin{subfigure}{0.6\textwidth}
        \includegraphics[width=\textwidth]{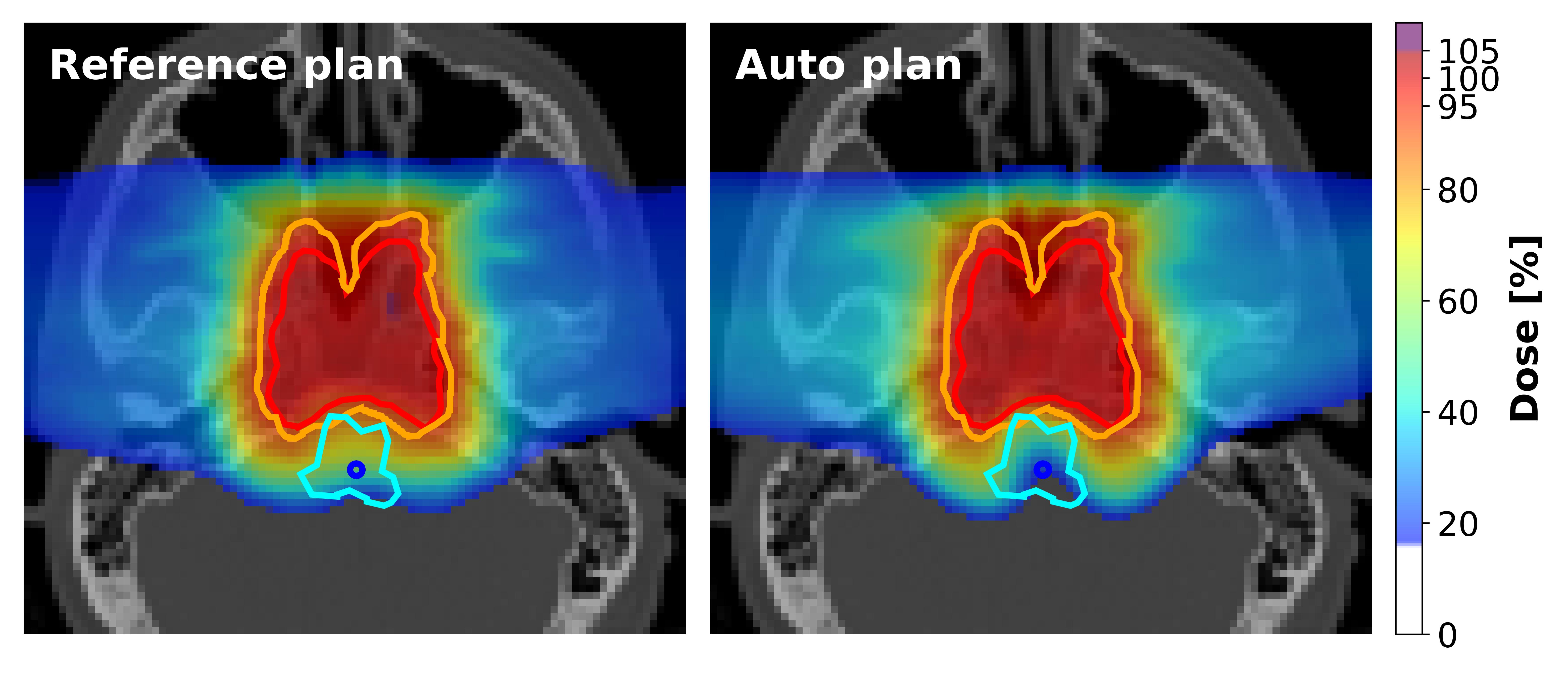}
        \caption{Dose distributions for patient 00. The left plot shows the reference plan, the right plot the auto plan. The color-wash dose-scale is displayed on the right-hand side of the figure}
    \end{subfigure}
    \caption{Results for patient 00. The colors of the contours correspond to the colors in the DVH plot.}
    \label{fig:00}
\end{figure}

\begin{figure}
    \centering
    \begin{subfigure}{0.6\textwidth}
        \includegraphics[width=\textwidth]{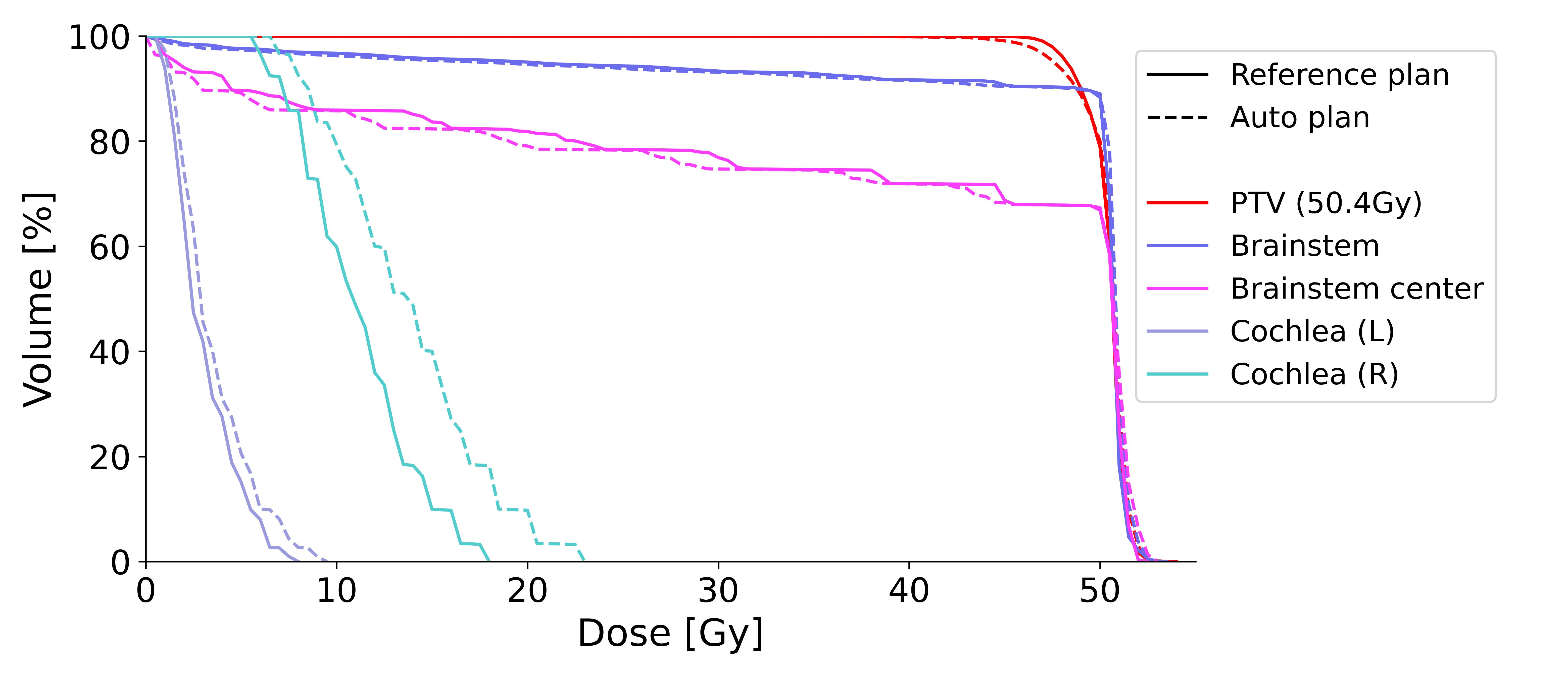}
        \caption{DVH curves for patient 01. The solid lines are the human-crafted reference plans, the dashed lines represent the auto plan.}
    \end{subfigure}\\
    \begin{subfigure}{0.6\textwidth}
        \includegraphics[width=\textwidth]{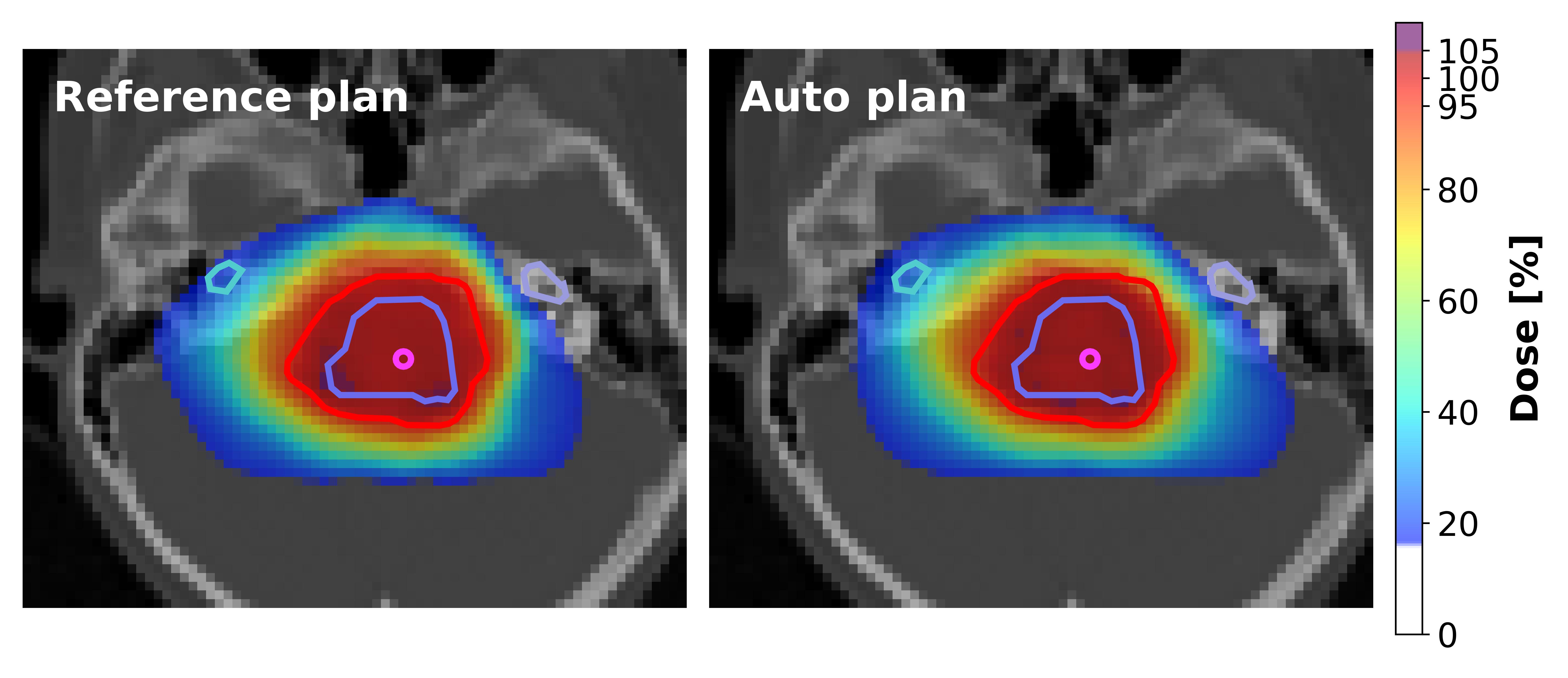}
        \caption{Dose distributions for patient 01. The left plot is the reference plan, the right plot the auto plan. The color-wash dose-scale is displayed on the right-hand side of the figure}
    \end{subfigure}
    \caption{Results for patient 01. The colors of the contours correspond to the colors in the DVH plot.}
    \label{fig:01}
\end{figure}

\begin{figure}
    \centering
    \begin{subfigure}{0.6\textwidth}
        \includegraphics[width=\textwidth]{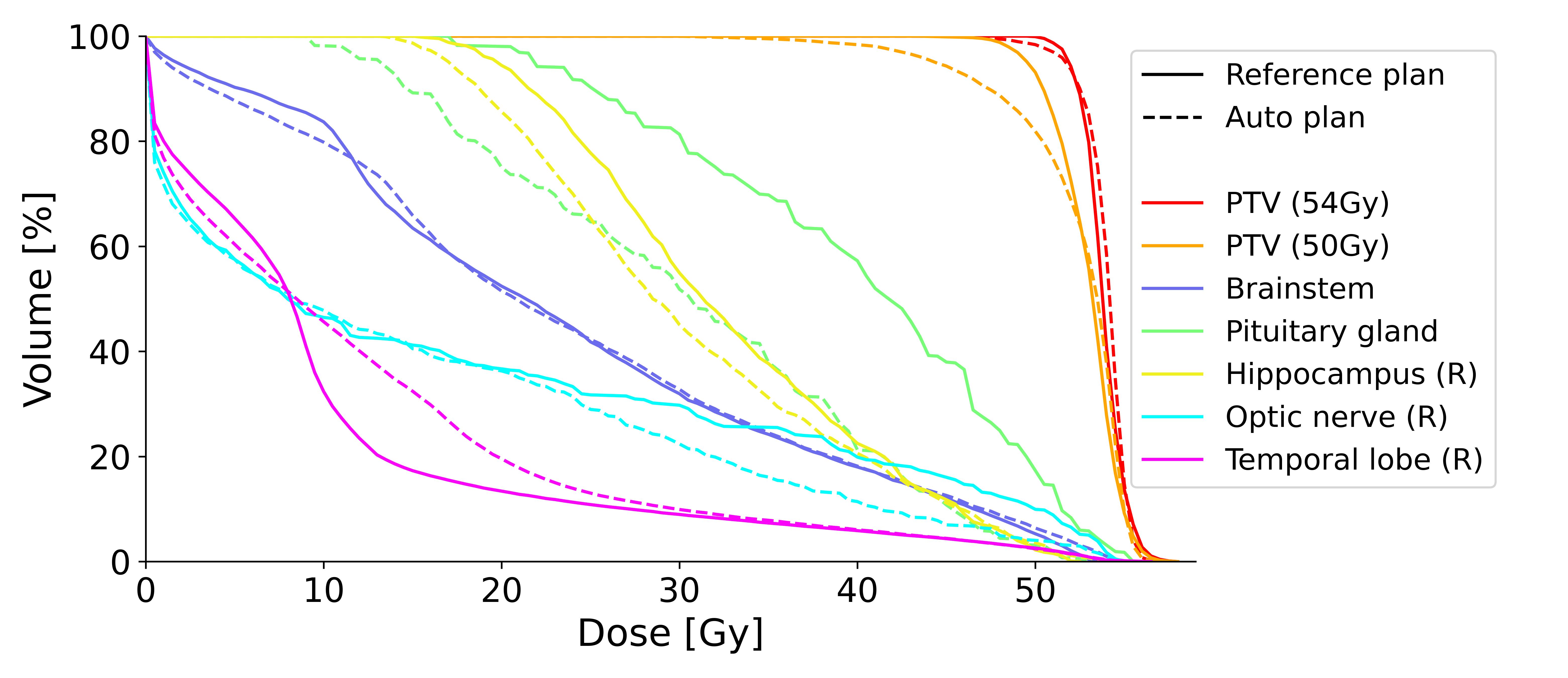}
        \caption{DVH curves for patient 04. The solid lines are the human-crafted reference plans, the dashed lines represent the auto plan.}
    \end{subfigure}\\
    \begin{subfigure}{0.6\textwidth}
        \includegraphics[width=\textwidth]{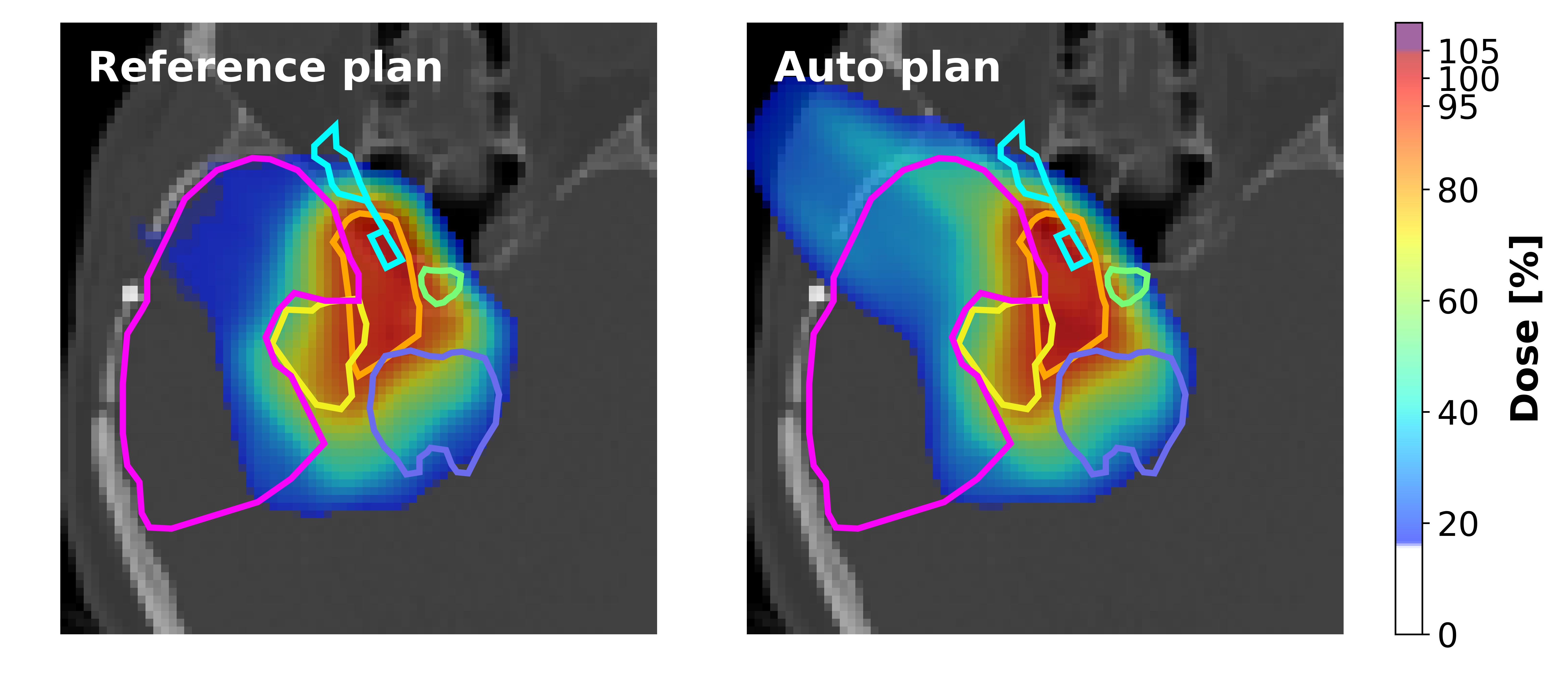}
        \caption{Dose distributions for patient 04. The left plot is the reference plan, the right plot the auto plan. The color-wash dose-scale is displayed on the right-hand side of the figure}
    \end{subfigure}
    \caption{Results for patient 04. The colors of the contours correspond to the colors in the DVH plot.}
    \label{fig:04}
\end{figure}

Three case studies are presented to showcase the similarity of the reference and auto plans. Patient 00 (see Fig.~\ref{fig:00}) is a case where the auto plan is considered superior. The dose distribution within the hotter target is more homogeneous ($\Delta HI = HI_{reference} - HI_{auto} \approx 2.9$), which can also be seen in the DVH curves (Fig.~\ref{fig:00_dvh}). The maximum dose in the entire dose distribution is reduced ($\Delta D_{max} = 0.3\%$). Most importantly, the auto plan is closer to fulfilling the brainstem constraint $D_{max, brainstem} \leq \SI{54}{Gy}$ ($D_{max} \approx \SI{61.1}{Gy}$ for the auto plan, $D_{max} \approx \SI{64.1}{Gy}$ for the reference plan). The center of the brainstem $D_{max}$ is reduced from \SI{50.7}{Gy} to \SI{37.8}{Gy}, which is far below the threshold value of $\SI{50.4}{Gy}$. On the other hand, the dose to normal tissue is increased in the auto plan, but not to an extent that would make the plan inacceptable. Patient 01 (Fig.~\ref{fig:01}) was considered unacceptable due to underdosage of the target volume in the right-anterior corner. Also for this patient, the dose to the normal tissue seems to be increased, which results in a higher dose to the cochleae. The dose to the brainstem and the hotspots within the target are comparable for both plans. Patient 04 (Fig.~\ref{fig:04}) is an example of a patient for which the reference plan and the auto plan are of comparable quality without one being superior to the other. As can be seen in Tab.~\ref{tab:metrics}, the auto plan has both a lower maximum dose ($\Delta D_{max} \approx 1.4\%$ of the prescribed dose) and a more homogeneous dose distribution ($\Delta HI \approx 0.18$) in the higher dose target. On the other hand, the auto plan has a lower minimum dose ($\Delta D_{min} \approx 9.7\%$) in the higher dose target and a lower $D_{95\%}$ ($\Delta V_{95\%} \approx 1.6\%$). The integral doses to the pituitary gland, right hippocampus and right optic nerve are reduced. The DVH curves of the right hippocampus and the brainstem cross for the reference plan and the auto plan, which prohibits a clear ranking of the plans w.~r.~t.\ these OARs. The entry-path dose is increased for the auto plan.

\begin{figure}
    \centering
    \includegraphics[width=0.5\textwidth]{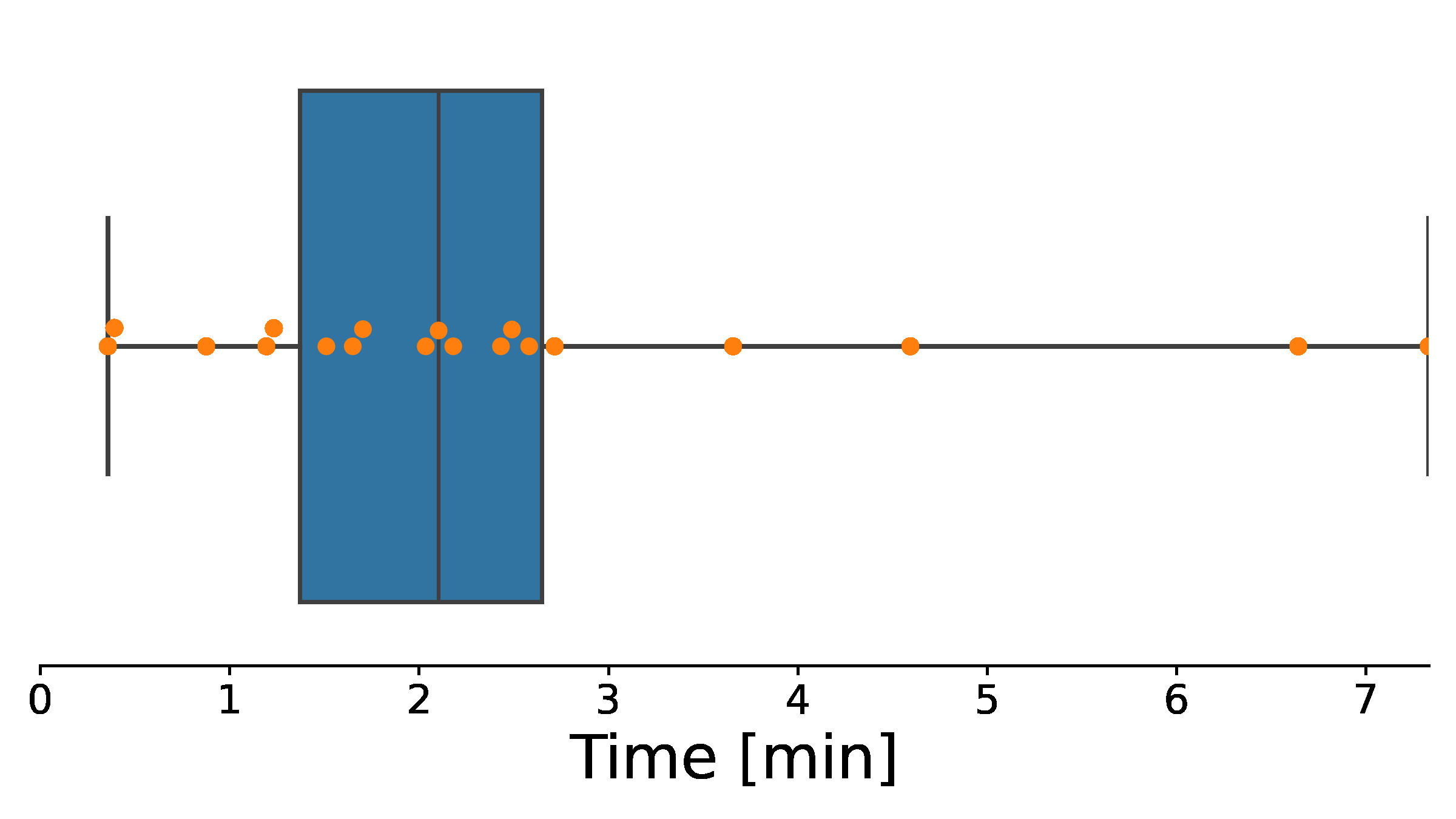}
    \caption{Time needed for the spot optimisation using JulianA. Each of the orange dots represents one patient. The boxes denote the lower/upper quartiles, the vertical line represents the median and the whiskers are the minimum and maximum values, respectively.}
    \label{fig:timing}
\end{figure}

The time needed for the generation of each plan is presented in Fig.~\ref{fig:timing}. These runtimes assume that the beam arrangement is given. Runtime for just-in-time (JIT) compilation and data loading is not included. This is a realistic estimate for an implementation in a treatment planning system (TPS). The data needs to be loaded regardless of whether an optimisation is performed. JIT compilation happens at every start of the program, but since the spot weight optimisations in our in-house TPS is performed on a computation server, that compilation time is only needed once at the start of the server, but not during the user interaction. The runtimes between $\SI{0.5}{min}-\SI{7}{min}$ with a median of $\SI{2}{min}$ (achieved on an NVIDIA DGX A100 GPU) is competitive with runtimes in the literature (Tab.~\ref{tab:literature_summary}).


The new loss function (Equ.~\ref{equ:loss1}) is a mathematical representation of the clinical intent and the objective weights in Tab.~\ref{tab:tradeoff_weights} introduce an interpretable "currency conversion". For example, $\alpha_{max} = 10$ and $\alpha_{oar, mean} = 1$ mean that a single voxel exceeding the maximum dose threshold by \SI{0.1}{Gy} is equivalent to exceeding the mean dose constraint of a single OAR by $\SI{1}{Gy}$. Such a conversion rate allows quantitative reasoning about planning tradeoffs, which has potential applications in statistical modelling and machine learning using historic treatment plans. Additionally, the loss function provides a plan score that quantifies plan desirability, therefore introducing a ranking between plans. Such a ranking has a broad range of applications like beam angle optimisation and machine learning. We intend to explore some of these paths in future work.

\section{Discussion}
JulianA aims to fill a niche that has not been covered in the literature so far. The existing automatic planning solutions are either slow (such as ECHO \cite{echo}, multi-objective optimisation methods \cite{sayed} or POPS \cite{pops}) or require additional patient-specific input such as objective weights \cite{matrad} or upper and lower bounds to the dose for each voxel \cite{projection}. On the other hand, JulianA does not require patient-specific parameters or objective prioritisation, is fast to execute, flexible and can be easily extended without increasing the runtime of the algorithm exceedingly.

Auto-planning for proton therapy utilising JulianA shows promising first results. The target coverage and OAR sparing are similar enough to the performance of a human-generated plan that an independent reviewer was not able to distinguish reliably between a machine-generated and a human-generated plan. Even though the new optimiser relies on objective weights, it can be considered an automatic method because the default weights can be used for a broad range of intra- and extracerebral neoplasms within the cranial region, as proven by the results on our dataset. As such, the default weights, although being site-specific are indication-agnostic. Further research is needed to investigate this hypothesis. Once the patient independent objective weights are chosen, no human interaction is needed, which renders JulianA a fully automatic planning method for the site studied here. Planning times between $\SI{0.5}{min}-\SI{7}{min}$ (mean and standard deviation $\SI{2.5}{min} \pm \SI{1.8}{min}$) are comparable if not better than previously published results (Tab.~\ref{tab:literature_summary}). It must be stated that this comparison is biased because of the vastly different hardware used in each study.

JulianA is a promising tool for both the clinic and research. It is written in the easily accessible and free and open source programming language Julia, which allows trivial scripting. The system is currently not implemented into our treatment planning system. Work is ongoing to export the resulting plans as DICOM RT plans, which will allow them to be imported into any TPS that can import DICOM RT plans. At the moment, dose distributions can be optimised and calculated using a standalone research tool. Machine parameters and phase space information are stored as structured configuration files that could be adjusted to match any treatment facility. Therefore, JulianA is easily available to domain experts who lack extensive programming skills. These experts have now a valuable tool to quickly generate high-quality treatment plans. On the one hand, researchers do not have to optimise a plan by hand to obtain a realistic treatment plan, which brings research closer to clinical practice. On the other hand, clinicians obtain a baseline plan that can be used for quality assurance or as a starting point for a manually crafted plan. Given that $16$ out of $19$ treatment plans were considered acceptable, the work load on human planners could be substantially reduced. At the same time, the convergence of each planning goal can be assessed, which reveals the bottle neck of a plan to a human planner and therefore facilitates the fine tuning of objective weights for especially challenging cases.

Both the algorithm and its implementation are flexible and can be trivially extended to incorporate additional planning goals. All that is needed is adding an additional term to the loss function. Even the gradient calculation is automated thanks to automatic differentiation, which does not require specialised programming skills. Since gradient-based algorithms do not suffer from the curse of dimensionality, adding additional loss terms does not increase the runtime substantially, which makes extension not only easy but also efficient, especially compared to other planning tools that are based on linear or non-linear programming or multi-objective optimisation.

There are many opportunities for extensions and applications. For instance, biological information such as tumour control probability and normal tissue complication probability models could be incorporated into the loss function. Further, the automatic planning method could be used to create consistent plan databases for machine learning applications (e.\ g.\ dose prediction models) or data mining that do not suffer from inter-planner variability. Future work should investigate the performance of JulianA for patients at other treatment sites. Given the promising results for the diverse set of patients in this pilot study, we expect the algorithm to work without changes except adjustments to the tradeoff parameters in Tab.~\ref{tab:tradeoff_weights}.

This work is only a proof-of-concept and future work is needed to incorporate JulianA into clinical practice. First, the dose in the normal tissue should be reduced by investigating different initialisation schemes. Second, a clever choice of optimisation points could reduce the planning time even further. Third, robustness w.\ r.\ t.\ CT calibration uncertainties and positioning errors has not been considered so far, but could be added as either an intermediate step when calculating the influence matrix or by calculating the loss terms in different scenarios and using the worst scenario to build the total loss function. Finally, further research should test JulianA for a larger patient cohort, a larger number of radiation oncologists for plan quality assessment and an even more detailed analysis of the strengths and shortcomings of JulianA. Especially, followup studies with patient cohorts of a single indication are needed to proof the utility of JulianA with stronger statistics. Further research should also investigate the robustness of the plans generated by JulianA.

\section{Conclusions}
The new spot weight optimisation algorithm JulianA has been presented and evaluated by a radiation oncologist. For this limited number of patients, the automatically generated plans were not distinguishable from the human generated plans in a majority of cases, even though they were not Pareto-optimal. The auto plans were created in only $\SI{0.5}{min} - \SI{7}{min}$ and fulfilled the clinical acceptability criteria for $16$ out of $19$ patients. JulianA has the potential to reduce the workload to human planners and bring realistic treatment plans into the reach of researchers without special treatment planning training. The code is implemented in the Julia programming language, which is powerful, easy to learn and fast even for domain experts without extensive programming skills. We hope that this work brings the treatment planning research closer to and facilitates quicker knowledge transfer into the clinic.

\section{Acknowledgements}

This work has been done as part of the INSPIRE project and is supported by a PSI CROSS project.

We acknowledge the assistance of Derek Feichtinger and
Marc Caubet for their help with the Merlin and Gwendolen cluster, which enabled the computational work of the research.

\section{Conflict of Interest Statement}
The authors have no relevant conflicts of interest to disclose.

\clearpage
\bibliographystyle{medphy}
\bibliography{refs}

\appendix
\section{Code and data availability}
The code is published on Zenodo~\cite{julianaCode}. The data are available upon request.

\section{Dose metrics}
\begin{table}
    \centering
    \begin{tabular}{rrrrrrrrrrr}
        \toprule
          & \multicolumn{2}{r}{$D_{min}$ [\%]} & \multicolumn{2}{r}{$V_{95\%}$ [\%]} & \multicolumn{2}{r}{$D_{max}$ [\%]} & \multicolumn{2}{r}{$HI$} & \multicolumn{2}{r}{$CI$}\\
        \cmidrule(rr){2-3} \cmidrule(rr){2-3} \cmidrule(rr){4-5} \cmidrule(rr){6-7} \cmidrule(rr){8-9} \cmidrule(rr){10 - 11}
          & reference & auto & reference & auto & reference & auto & reference & auto & reference & auto\\
        \midrule
        patient 00 & 70.2 & 68.9 & 91.4 & 95.3 & 105.4 & 105.1 &  8.3 &  5.4 &  1.49 & 1.42\\
        patient 01 & 89.3 & 73.9 & 96.8 & 94.1 & 106.2 & 105.5 &  3.3 &  4.2 &  0.99 & 0.97\\
        patient 02 & 80.0 & 65.7 & 81.4 & 83.8 & 111.4 & 106.4 &  3.2 &  3.7 &  0.90 & 0.93\\
        patient 03 & 78.5 & 81.6 & 93.9 & 91.8 & 106.7 & 105.1 &  6.1 &  6.7 &  1.21 & 0.88\\
        patient 04 & 92.2 & 82.6 & 98.0 & 96.4 & 107.0 & 105.6 &  3.8 &  3.7 &   1.27 & 1.12\\
        patient 05 & 77.2 & 68.6 & 89.0 & 88.1 & 108.9 & 106.2 &  5.8 &  7.1 &  0.95 & 1.00\\
        patient 06 & 89.1 & 65.8 & 96.2 & 88.4 & 104.8 & 105.8 &  3.6 &  6.7 &  1.02 & 1.01\\
        patient 07 & 64.9 & 52.5 & 86.7 & 90.1 & 110.2 & 105.6 & 13.6 & 11.0 &  1.35 & 1.32\\
        patient 08 & 79.0 & 66.1 & 93.4 & 96.1 & 109.8 & 105.8 &  5.7 &  3.9 &  0.95 & 1.01\\
        patient 09 & 56.2 & 75.1 & 86.2 & 89.6 & 105.5 & 105.6 & 12.8 &  9.8 &  1.49 & 1.49\\
        patient 10 & 64.7 & 59.5 & 98.2 & 94.6 & 106.6 & 105.3 &  1.4 &  2.1 &  1.02 & 1.07\\
        patient 11 & 89.4 & 77.0 & 95.2 & 94.2 & 106.0 & 105.5 &  4.9 &  5.0 &  0.97 & 0.98\\
        patient 12 & 79.1 & 88.2 & 94.1 & 97.8 & 105.2 & 105.6 &  3.8 &  3.1 &  1.57 & 1.42\\
        patient 13 & 76.5 & 53.4 & 96.0 & 95.1 & 110.0 & 105.9 &  4.4 &  4.5 &  0.99 & 1.06\\
        patient 14 & 86.7 & 72.5 & 89.7 & 81.7 & 108.3 & 107.0 &  5.7 & 10.0 &  1.05 & 1.12\\
        patient 15 & 37.7 & 63.1 & 78.4 & 81.9 & 110.5 & 106.5 & 25.3 & 15.2 &  1.53 & 1.48\\
        patient 16 & 75.6 & 87.6 & 96.3 & 98.2 & 104.2 & 104.9 & 3.5  &  3.4 & 3.80 & 2.99\\
        patient 17 & 76.0 & 72.2 & 96.4 & 93.0 & 106.9 & 105.1 & 4.6  &  5.8 &  0.99 & 0.99\\
        patient 18 & 88.7 & 58.4 & 95.9 & 83.3 & 107.0 & 105.6 & 4.4  & 10.9 &  0.98 & 0.95\\
        \bottomrule
    \end{tabular}
    \caption{Dose metrics for the reference and the auto plan for each patient. The minimum dose $D_{min}$, $V_{95\%}$ and $HI = D_{5\%} - D_{95\%}$ are calculated for the PTV. If multiple PTVs are present, the one receiving the highest dose is selected. The $D_{max}$ is the overall maximum dose in any voxel. The conformity index CI is the ratio between the number of voxels that receive at least $95\%$ of the prescribed dose (lowest dose if there are multiple dose levels) to the number of voxels in the target with the lowest-dose prescription.}
    \label{tab:metrics}
\end{table}

\begin{figure}
    \centering
    \includegraphics[width=\textwidth]{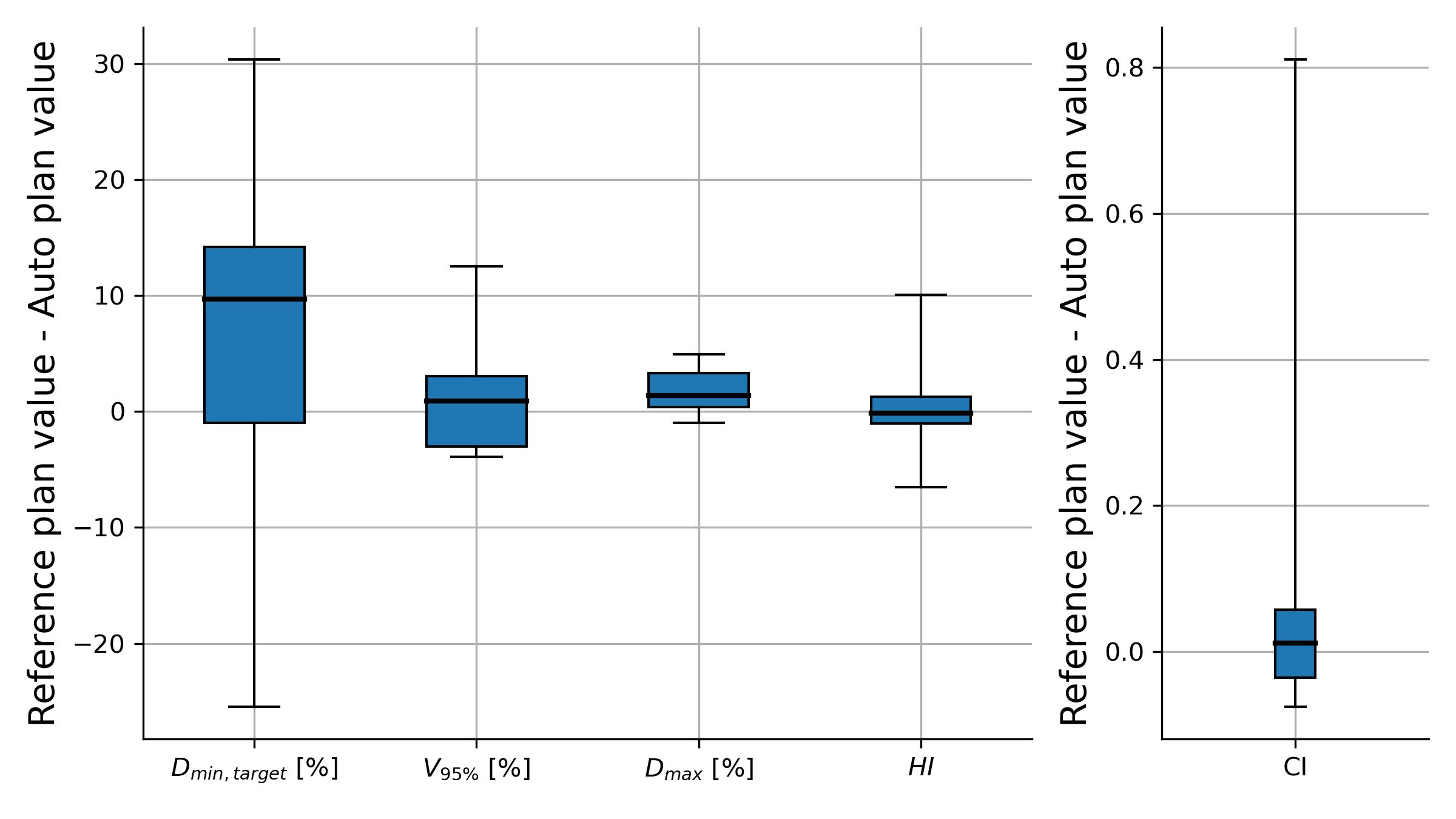}
    \caption{Boxplots for the autoplanning values in Tab.~\ref{tab:metrics}.}
    \label{fig:boxplots}
\end{figure}

\end{document}